\documentclass[final,5p,times,twocolumn]{elsarticle}

\usepackage{amssymb}
\usepackage{amsmath}
\usepackage{graphicx}
\usepackage{booktabs}
\usepackage{float}
\usepackage{fancyhdr}
\usepackage{makecell}
\usepackage[numbers]{natbib}
\usepackage{microtype}
\usepackage{multirow}
\usepackage{hyperref}
\usepackage{subcaption}
\usepackage{caption}
\usepackage{orcidlink}
\journal{}

\begin{document}

\begin{titlepage}
	\centering
	\vspace*{1cm}
	\Huge
	\textbf{Rice Price Dynamics during the 1945--1947 Famine in Post-War Taiwan: A Quantitative Reassessment}
	\vspace{0.5cm}
	
	\LARGE
	\vspace{1.5cm}
	\textbf{Huai-de Chen\orcidlink{0009-0005-3339-2797}\footnote{Email: 2519010352@sxnu.edu.cn} and Hai-liang Yang\footnote{Corresponding author: Hai-liang Yang, Email: 703167@sxnu.edu.cn}}
	\vspace{1cm}
	
	\Large
	\textbf{School of Marxism, Shanxi Normal University} \\
	\textbf{No. 339, Taiyu Road, Xiaodian District, Taiyuan, Shanxi 030031, China}\\
	\vfill
	\large
	\textbf{Submission Date: 2025/12/02}
\end{titlepage}


\begin{frontmatter}



\title{Rice Price Dynamics during the 1945--1947 Famine in Post-War Taiwan: A Quantitative Reassessment}


\author[university]{Huai-de Chen\orcidlink{0009-0005-3339-2797}}
\author[university]{Hai-liang Yang\corref{cor1}}
\cortext[cor1]{Corresponding author}
\ead{703167@sxnu.edu.cn}
\address[university]{School of Marxism, Shanxi Normal University, Taiyuan, Shanxi 030031, China}
\begin{abstract}
We compiled the first high-frequency rice price panel for Taiwan from August 1945 to March 1947, during the transition from Japanese rule to China rule. Using regression models, we found that the pattern of rice price changes could be divided into four stages, each with distinct characteristics. Based on different stages, we combined the policies formulated by the Taiwan government at the time to demonstrate the correlation between rice prices and policies. The research results highlight the dominant role of policy systems in post-war food crises.
\end{abstract}

\begin{graphicalabstract}
\includegraphics[width=1\textwidth]{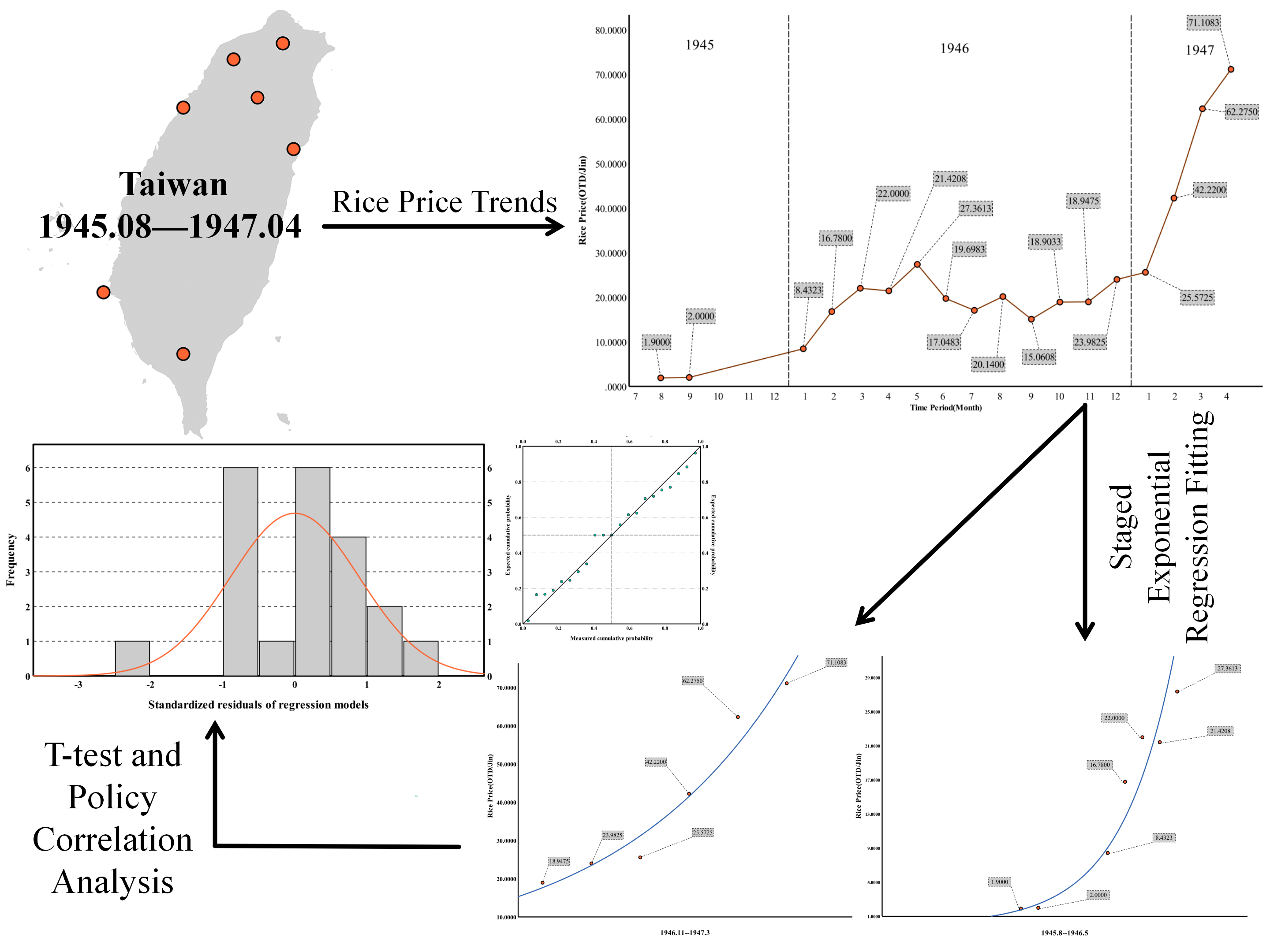}
\end{graphicalabstract}

\begin{highlights}
\item The first high-frequency rice price panel for Taiwan from August 1945 to March 1947
\item Four stages describing the pattern of rice price trend
\item Highlighting the dominant role of policy systems in post-war food crises
\end{highlights}


\end{frontmatter}


\section{Introduction}
\subsection{Historical Background: The End of the Second Sino-Japanese War}
We started narrating the historical background from 1943. On December 1st, 1943, the heads of state of the United States, China, and the United Kingdom jointly issued the Cairo Declaration, which stipulated the ownership of Taiwan after World War II: "It is their purpose that Japan shall be stripped of all the islands in the Pacific which she has seized or occupied since the beginning of the first World War in 1914, and that all the territories Japan has stolen from the Chinese, such as Manchuria, Formosa\footnote{Formosa is another name for Taiwan.}, and The Pescadores, shall be restored to the Republic of China."\cite{r1}Similarly, the Potsdam Proclamation also stipulated that Taiwan belonged to China after the war.

On August 15th, 1945, Japan surrendered unconditionally, and the sovereignty of Taiwan was legally returned to China. On September 1st of the same year, the Chinese Nanjing National Government established the Taiwan Provincial Garrison Command in Chongqing. On September 3rd, Chiang Kai-shek issued an order: "Appoint General Chen Yi as the chief official for accepting the surrender of Taiwan and the Penghu Islands."\cite{r2}At 10 a.m. on October 10th, all officers and soldiers of the Forward Command Post, along with local gentry and civilians from Taiwan, gathered at the former Guild Hall\footnote{It's now known as Taipei Zhongshan Hall.}in Taipeito hold the first National Day commemoration in Taiwan Province, declaring the return of Taiwan's sovereignty.
\subsection{Event Background: Food Shortage(1945.9-1947.3)}
In August 1945, the price of rice once dropped below 2 Old Taiwan dollar(OTD\footnote{Old Taiwan dollar (1946-1949), predecessor to today’s New Taiwan dollar, withdrawn at 40,000:1 owing to post-war hyperinflation. We abbreviate the Old Taiwan dollar as OTD.}) per jin\footnote{Jin is a traditional chinese unit of measurement. Commonly, one jin is equal to 500 grams.}. From mid-August to mid-September 1945, the market in Taiwan Province once presented a scene of abundant goods and a steady decline in prices. However, From the second half of September in the 34th year of the Republic of China\footnote{September 1945.}, the price of grain soared. In early 1947, the price of rice in Taipei rose to 14 OTD per jin. Later, under the influence of the gold rush in Shanghai, the price of rice suddenly jumped to over 30 OTD per jin, and even reached more than 40 OTD per jin on the black market.

From this, we have a question: Can we quantify the fluctuations in rice prices during this period? Then, the deeper question is whether we can identify the factors related to the changes in grain prices?

\subsection{Financial Background: Currency Change(1945-1946)}
After Japan's surrender in 1945, Taiwan was liberated. The original currency of the Japanese colony - the Taiwan banknote - needed to be cleared and replaced. Starting from September 1, 1946, old Taiwan banknotes were redeemed at a ratio of 1:1. The original term was two months, ending on October 31. However, as only over 1.71 billion yuan was recovered on October 12th, accounting for 41 percents of the total circulation, in order to take into account the interests of the bondholders, with the approval of the Ministry of Finance by the Chief Executive's Office, the redemption period was extended by one month. The collection scope includes the old Taiwan banknotes in circulation in the market. These banknotes were the currency issued during the Japanese colonial period, and their circulation amount was approximately 3.9 billion yuan in 1946. The recipients of the exchange are the public and institutions holding old Taiwan banknotes. Old Taiwan banknotes and newly issued old New Taiwan dollars (New Taiwan dollar exchange certificates) can be exchanged at a ratio of 1:1. By the final deadline, a total of 3.44 billion yuan worth of Taiwan banknotes had been withdrawn, basically completing the clearance of old Taiwan banknotes. Taiwan's financial system has thus entered a new stage. All the old Taiwan banknotes retrieved were destroyed.

\section{Literature Review}
We learned from the information released by the official Taiwan government that at that time, there were mainly three types of rice sold in the Taiwan market, including superior rice or known as the top-quality rice, medium-quality rice and inferior-quality rice. To unify the research criteria, we only studied the price data of top-quality rice. At that time, the Chen Yi government would regularly compile Statistics on prices in Taiwan and publish the official statistics through the publication Taiwan Monthly of Commodity-price Statistics. Taiwan Monthly of Commodity-price Statistics is a statistical publication edited and distributed by the Statistics Office of the Taiwan Provincial Administrative Office in Taipei in 1946. Its publication cycle is monthly. It can be seen from each statistical item that the journal has taken into account the regional differences among Taipei City, other counties and cities in Taiwan Province, and major cities in the country, and attaches great importance to comparisons, making its data more valuable for reference. Fortunately, all the oil prints of this monthly magazine are included in the National Periodical Index System established by the Shanghai Library. For instance, in Appendix 2 of the 12th issue of Taiwan Monthly of Commodity-Price Statistics in 1946, the monthly and annual averages of rice prices within a certain period of time at that time were recorded in detail, i.e., In 1946, the official monthly average of rice prices in Taipei City in January was 8.2575 OTD per jin, the monthly average in December was 23.9825 OTD per jin, and the official annual average of rice prices in 1946 was 19.1334 OTD per jin\cite{r4}.

Some scholars have conducted research on the food shortage during this period. Many scholars have studied the rice price data at some specific time points. Zeng Lei-lei pointed out that the price of rice in Taiwwan rose above 2 OTD per jin on October 10th and soared to 8 OTD/jin by the end of December\cite{r7}. Weng Jia-hsi: "In January 1946, one jin of rice cost only 8.84 OTD."\cite{r8}On page 265 of Weng's paper, Weng presented a table of the prices of major daily necessities in Taipei City in the early post-war period. Among them, the price of rice was 8.84 OTD/jin in January 1946 and rose to 42.67 OTD/jin in February 1947, with an increase multiple of 4.83\cite{r8}. Professor Wang Xiaobo from Taiwan believes that according to the investigation data of some scholars, the price of rice in Taiwan was 6.3 OTD per jin in January 1946 and 32.33 OTD per jin in February 1947. However, we remain skeptical about the data provided by Wang Xiaobo because the gap between the data he provided and those provided by other scholars exceeds 23 percents, and it is even larger than the official data of the Taiwan government. We fully doubted that the data he provided are for other types of rice, not for top-quality rice.

Additionally, we have found many official documentary books that record the food shortage events that occurred in Taiwan during this period. The book "Taiwan's Recovery and the Provincial Conditions of the Five Years After Recovery", edited by Chen Mingzhong and Chen Xingtang and published by Nanjing Publishing House, includes many reports from newspapers and periodicals during the grain shortage incident. We believe that the representative newspaper of these newspaper types is Wenhui Daily. These two individuals are researchers at the Second National Historical Archives of China, as we believe the content of this book is highly credible\cite{r5}. Given that this material has not been reprinted since 2008 and it cannot be found in the bookstore, to Science Data Bank\cite{r6} we uploaded the scanned copy whose filename is RefBook1.pdf\footnote{DOI: 10.57760/sciencedb.32762}. In March 1947, several reports in Wenhui Daily mentioned the price of rice in the Taiwanese market at that time. Among them, the reports of the two journalists caught our attention. Journalist Feng Yan's report mentioned that the price of rice was 13.5 OTD per jin in January 1947, soon rose to 26 OTD per jin, and by early March, it had reached around 30 OTD per jin. Journalist Yang Feng left Taiwan before his article was published in Wenhui Daily on March 4, 1947. Yang Feng said that the price of rice at that time was around 40 OTD per jin. Based on this, it can be inferred that the figure "42.67 OTD per jin" given by scholar Weng Jiaxi might have existed in late February 1947.

We studied the calculation method of the official monthly price report data in Taiwan and found that the average monthly rice price is derived from the weighted average of the prices on the 5th, 15th and 25th of each month. The data provided by scholars differs from the official data. After eliminating the obviously unreasonable data, some data still have an error of less than 7 percents. However, at present, we have not found a systematic organization and quantitative analysis of these data. It would be a pity if these data were not further studied but merely sealed in historical records.

Hayek believed that the harm brought by inflation to society is serious. People generally summarize these harms as rising prices, social unrest and people's unease, etc. Therefore, we believe that the fluctuations in rice prices in Taiwan after the war may be related to government policies and some major events. “Inflation is not merely a matter of rising prices; it is also a process that distorts relative prices, undermines the coordinating function of the market, and creates social unrest.”\cite{r9} Some policies of the government have not effectively curbed inflation, which has led to the increase in prices, including the price of rice. Therefore, we need to conduct further research on the correlation between policies and rice prices.
\section{Research Methods}
First of all, the most fundamental work we need to do is to sort out the original monthly average data of rice prices with credibility. The monthly average data of rice prices in the internal price report of Chen Yi government is relatively complete and authoritative, and it is the data we first accept. Of course, we believe that this data may have a tendency to be glorified at certain times, so we need to introduce other data to correct it. We simultaneously adopted all the data from Weng's paper and Zeng's paper.

Then, we performed a weighted average of different data from the same month. Chen Huai-de believed that official data should be regarded as the main source of trust for data, and its proportion in the weighted average calculation should not be less than 50 percents. After discussion, we finally determined that the weight of the official monthly price report data in the weighted average calculation is 0.7, and the total weight of other data is 0.3. For instance, when estimating the rice price in January 1946, we adopted a weighted average method of 8.84 as mentioned by Weng Jiaxi and 8.2575 as the "official data". We accepted 70 percents of the official data and 30 percents of what Weng Jiaxi said. We stipulated that the error should be within 5 percents, and thus the revised average monthly rice price for that month was 8.4323±0.4 OTD per jin. By analogy, we obtained the monthly average revised value of rice prices in the Taipei market from August 1945 to March 1947.

There is still one doubt that we didn't answer just now. Why would we rather sacrifice the sample size to perform a weighted average on these data and obtain the corrected value? The primary reason is that the official data itself is calculated using weighted averages, while scholars usually have data at one or several time points. If we don't do this, when we perform some operations on these data later, such as building a regression model for them, a series of problems will arise, for instance, the residual results of the regression model will be very poor. The secondary reason is that these data are not 100 percents reliable. Even the official monthly price report may be at risk of being tampered with or beautified. Based on the timeline, we conduct data screening, eliminating a small amount of obviously outrageous data, and then perform weighted averaging on different data of the same month. This can statistically minimize the problem of individual case errors as much as possible. For the missing data of certain months, we adopt some estimates based on vague narratives from historical materials, or we omit the unimportant parts. We compiled the corrected Data into an SPSS dataset and uploaded it to Science Data Bank\cite{r3}.

Then, we used tools such as SPSS to conduct fundamental quantitative analysis and advanced in-depth quantitative analysis on these corrected data. For instance, when conducting fundamental quantitative analysis, Yang Hai-liang proposed a method that involves segmenting the data and then regression fitting. To conduct reasonable data segmentation, we first need to judge the general changing trends of these data, where we can use the line chart method for visual presentation.

There are many methods that can be adopted for advanced quantitative analysis. The first thing that comes to our mind is the t-test. We can introduce a 0-1 variable for a t-test. 0-1 programming is a type of integer programming where the decision variable is only 0 or 1, and it falls within the category of combinatorial optimization mathematical models. We assigned all the policy variables at the sampling points after the Chen Yi government announced in October 1946 that the public was required to exchange the currency in their hands for Old Taiwan dollars within 60 days to 1, and all the variables at the sampling points before that to 0. Based on this processing, we conducted a t-test on these data in spss. Next, we postpone the inflection points of the policy variables 0-1 by different months and introduce the lag1, lag2, and lag3 variables. Based on this, we constructed a regression model for further verification.
\section{Results and Analysis}
\subsection{Preliminary Quantitative Analysis}
\begin{figure}
	\centering
	\includegraphics[width=0.48 \textwidth]{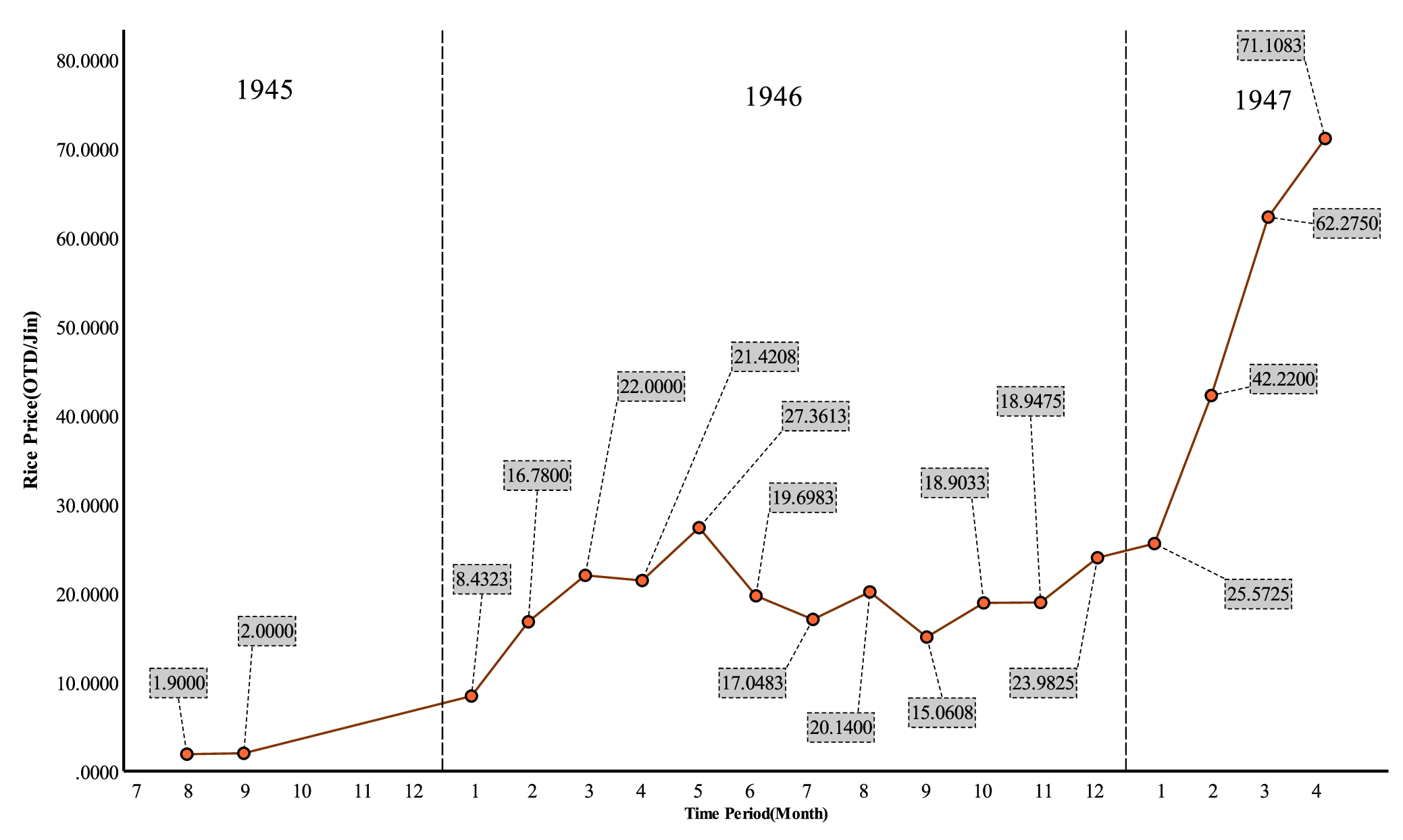}
	\caption{Trend of monthly average rice price (1945.8--1947.3).}
	\label{fig:1}
\end{figure}

The trend of rice prices during the grain shortage from August 1945 to March 1947 is shown in Figure.~\ref{fig:1}.
As far as we're concerned, based on the the Figure.~\ref{fig:1}, grain shortage can be divided into four stages. The first stage of the food shortage was from August 1945 to September 1945. During this period, Taiwan was in the early stage of post-war recovery. There was no shortage of food, and prices were generally stable with a slight decline. The second stage was from September 1945 to May 1946. This stage was the first peak period of the grain shortage, during which grain prices rose relatively seriously. The third stage was from May 1946 to November 1946. This stage was in the low point of grain shortage, with grain prices remaining relatively stable and market prices showing a steady downward trend. The fourth stage was from November 1946 to April 1947. This stage was the second peak period of the grain shortage. Grain prices fluctuated sharply during this period, but the overall grain prices remained sharp for a long time. The average price in Taiwan more than doubled compared to the third stage, and social unrest occurred in Taiwan. After May 1947, grain prices showed a downward trend. The price of rice once dropped below 40 yuan per jin.

Due to the default data, there is less data in the first stage. We might as well combine the first and second stages into one large part. Then, we established regression models for the data of the three parts respectively.The notable feature of rice prices during the third stage of the grain shortage was the volatile decline in market prices. I believe that the data from this period did not have a clear co-directional trend or economic logic. Therefore, no regression modeling was conducted on the rice price data from May 1946 to November 1946.

At first, we attempted linear regression but failed due to the low degree of fit. Then, Yang Hai-liang proposed the method of fitting quadratic or cubic curves. However, although the model parameters of the two regression models we obtained were better than those of the first attempt, Chen Huaide found that no matter whether quadratic or cubic functions were used as the template functions of the regression models, there was a common problem, that is, there were trends that did not conform to the facts or fit values that were far from the actual values at both ends or in the middle of the fitting curve. After numerous experiments, we finally adopted the exponential function as the template function of the regression model.

\begin{figure}[t]
	\centering
	\begin{subfigure}{0.45\textwidth}
		\centering
		\includegraphics[width=\textwidth]{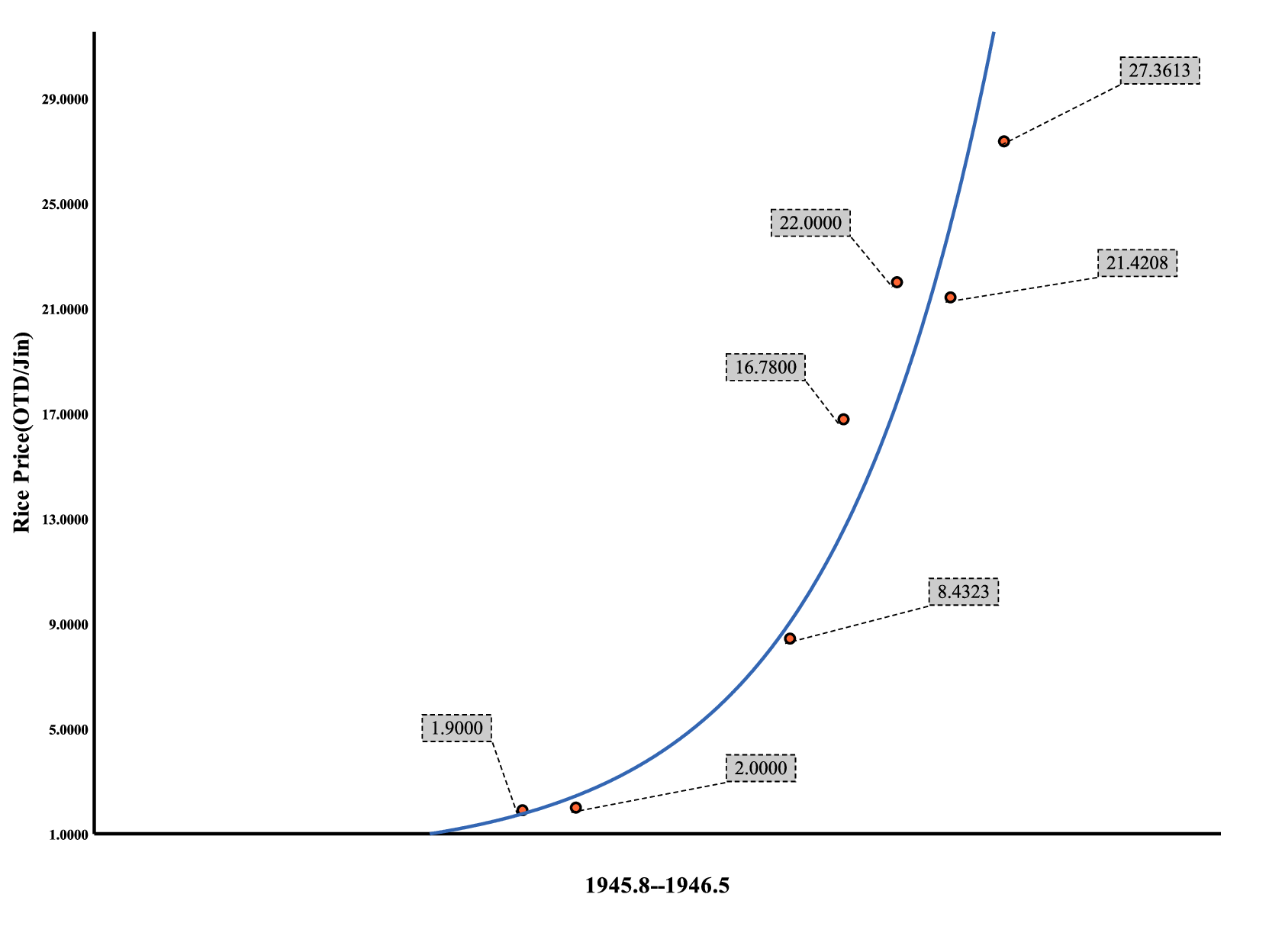}
		\caption{Fitting diagram of the rice price fluctuation model in the first two stages of the grain shortage from August 1945 to May 1946.}
		\label{fig:2}
	\end{subfigure}
	\hfill 
	\begin{subfigure}{0.45\textwidth}
		\centering
		\includegraphics[width=\textwidth]{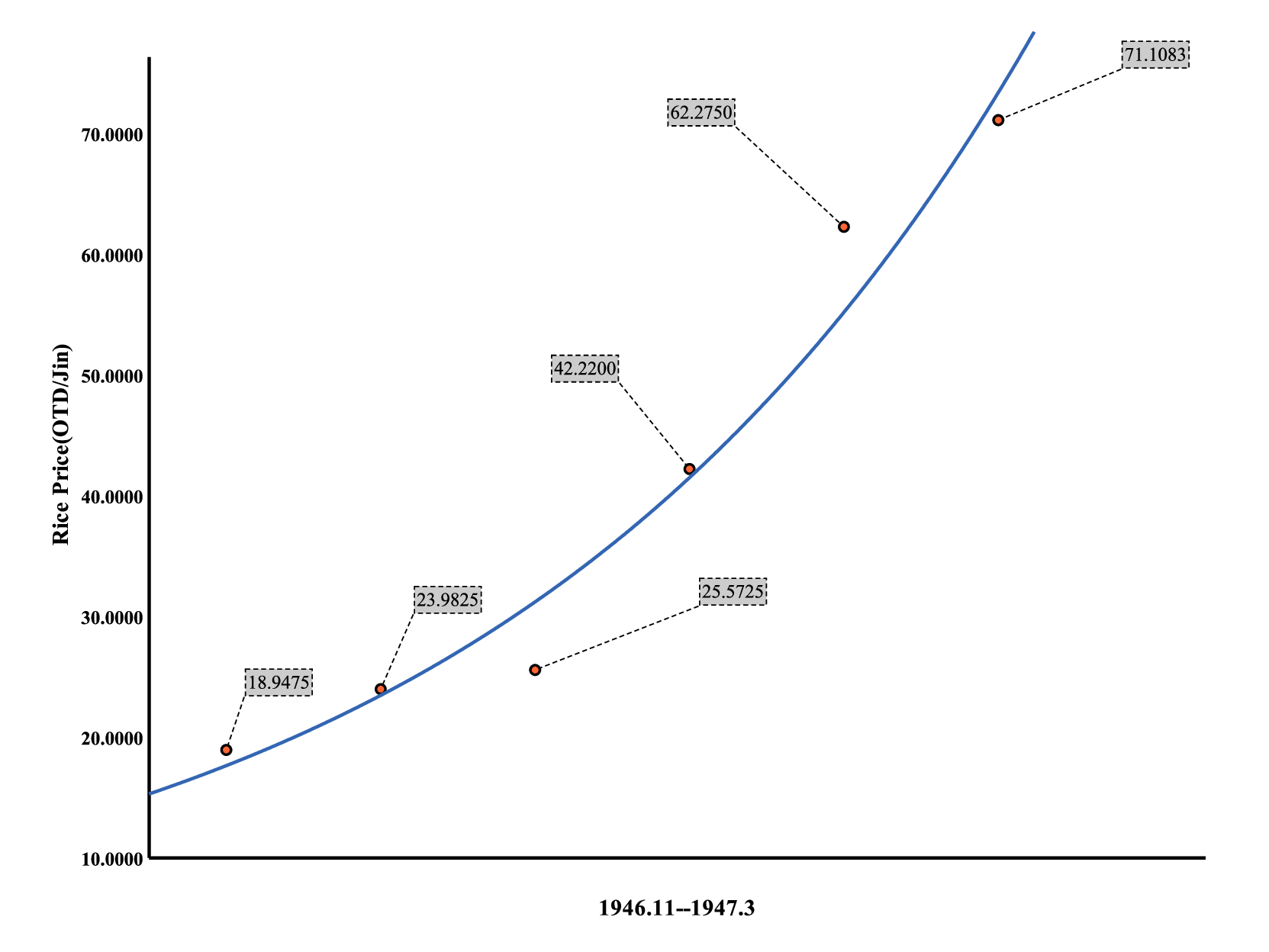}
		\caption{Fitting diagram of the rice price fluctuation model in the fourth stage of the grain shortage from November 1946 to March 1947.}
		\label{fig:3}
	\end{subfigure}
	\caption{}
	\label{fig:figure2}
\end{figure}

The results are as presented in Figure.~\ref{fig:2} and Figure.~\ref{fig:3} and Table.~\ref{tab:1} and Table.~\ref{tab:2}. We have obtained the most perfect quantitative results among several attempts. Subsequently, we sorted out the summary table of the regression model, drew the fitting curve graph of the regression model, and uploaded them to the Science Data Bank\cite{r3}.

\begin{table}[t]
	\centering
	\begin{minipage}[t]{0.48\textwidth} 
		\centering
		\setlength{\tabcolsep}{1pt} 
		\caption{Exponential model summary and parameter estimates for grain-shortage stages 1--2 (Aug.\ 1945--May.\ 1946).}
		\label{tab:1}
		\begin{tabular}{lccccccc}
			\toprule
			\multirow{2}{*}{Equation} &
			\multicolumn{5}{c}{Model summary} &
			\multicolumn{2}{c}{Parameter estimates} \\
			\cmidrule(lr){2-6}\cmidrule(lr){7-8}
			& $R^{2}$ & $F$  & df$_1$ & df$_2$ & $p$ & Constant & $b_1$ \\
			\midrule
			Exponential & 0.969 & 154.85 & 1 & 5 & $<$.001 & 0.128 & 0.327 \\
			\bottomrule
		\end{tabular}
	\end{minipage}
	
	\hfill 
	
	\begin{minipage}[t]{0.48\textwidth} 
		\centering
		\setlength{\tabcolsep}{1pt} 
		\caption{Exponential model summary and parameter estimates for Stage 4 of the grain shortage (Nov.\ 1946--Mar.\ 1947).}
		\label{tab:2}
		\begin{tabular}{lccccccc}
			\toprule
			\multirow{2}{*}{Equation} &
			\multicolumn{5}{c}{Model summary} &
			\multicolumn{2}{c}{Parameter estimates} \\
			\cmidrule(lr){2-6}\cmidrule(lr){7-8}
			& $R^{2}$ & $F$ & df$_1$ & df$_2$ & $p$ & Constant & $b_1$ \\
			\midrule
			Exponential & 0.959 & 93.13 & 1 & 4 & $<$.001 & 0.025 & 0.285 \\
			\bottomrule
		\end{tabular}
	\end{minipage}
	
	\hfill

	\begin{minipage}[t]{0.48\textwidth} 
		\centering
		\setlength{\tabcolsep}{1pt} 
		\caption{Descriptive Statistics}
		\label{tab:3}
		\begin{tabular}{lcccc}
			\toprule
			& Number & Averages & Standard deviation & Average error \\
			\midrule
			Policy var & & & & \\
			0 & 14 & 13.345843 & 8.5589676 & 2.2874803 \\
			1 & 7 & 37.572729 & 21.5274259 & 8.1366022 \\
			\bottomrule
		\end{tabular}
	\end{minipage}
\end{table}
\subsection{Advanced Quantitative Analysis}
Firstly, we conducted descriptive statistical analysis on the rice price of the two groups (defined by the Policy dummy variable "policy 0-1 var"). The statistical data of Group 0 and Group 1 are shown in Table.\ref{tab:3}.

Then, we conducted an independent sample t-test to assess whether there was a significant difference in the average rice price between the two groups, as shown in Table.\ref{tab:4}. In both cases (regardless of whether equal variance is assumed or not), the P-value of the t-test is less than 0.05, indicating a significant difference in the average rice price between the two groups. 

We also calculated several effect size indicators to assess the actual significance of the differences between the two groups, as shown in Table.\ref{tab:5}. Cohen's d: 14.02, 95 percents confidence interval [-1.73, -2.77] to [-0.65]. Hedges' g: 14.60, with a 95 percents confidence interval of [-1.66, -2.66] to [-0.63]. Glass's $\Delta$ : 21.53, with a 95 percents confidence interval of [-1.13, -2.20] to [0.01]. These effect size indicators indicate that the differences between the two groups are very significant and have practical significance. Therefore, we have sufficient reasons to believe that there is a significant difference in rice prices before and after the implementation of the currency exchange policy by the Chen Yi government, and the effect size indicator shows that this difference is of practical importance and can support the correlation between this policy and rice prices.
\begin{table*}
	\centering
	\setlength{\tabcolsep}{5pt} 
	\caption{Results of Independent Samples Test for Rice Price}
	\label{tab:4}
	\begin{tabular}{lcccccccc}
		\toprule
		& \multicolumn{2}{c}{\textbf{Levene's Test}} & \multicolumn{6}{c}{\textbf{t-Test for Equality of Means}} \\
		\cmidrule(lr){2-3} \cmidrule(lr){4-9}
		& F & Sig. & t & df & Sig. (2-tailed) & Mean Diff. & SE Diff. & 95\% CI \\
		\midrule
		Assume Equal Var. & 14.226 & 0.001 & -3.734 & 19 & 0.001 & -24.227 & 6.488 & (-37.807, -10.646) \\
		Not Assume Equal Var. & & & -2.866 & 6.966 & 0.024 & -24.227 & 8.452 & (-44.233, -4.221) \\
		\bottomrule
	\end{tabular}
\end{table*}
\begin{table*}
	\centering
	\setlength{\tabcolsep}{5pt} 
	\caption{Effect Sizes for Independent Samples Test on Rice Price}
	\label{tab:5}
	\begin{tabular}{lcccc}
		\toprule
		\textbf{Effect Size Measure} & \textbf{Standardized Measure} & \textbf{Point Estimate} & \textbf{95\% CI Lower} & \textbf{95\% CI Upper} \\
		\midrule
		Cohen's $d$ & Pooled SD & -1.728 & -2.771 & -0.653 \\
		Hedges' $g$ & Pooled SD with correction & -1.659 & -2.659 & -0.627 \\
		Glass' $\Delta$ & Control group SD & -1.125 & -2.195 & 0.005 \\
		\bottomrule
	\end{tabular}
\end{table*}
\begin{table*}
	\centering
	\caption{Descriptive Statistics}
	\setlength{\tabcolsep}{5pt} 
	\begin{tabular}{lccc}
		\hline
		Variable & Mean & Standard Deviation & Number of Cases \\
		\hline
		rice\_price & 21.421471 & 17.9888439 & 21 \\
		policy\_dummy & 0.33 & 0.483 & 21 \\
		policy\_lag1 & 0.29 & 0.463 & 21 \\
		policy\_lag2 & 0.24 & 0.436 & 21 \\
		lag3 & 0.19 & 0.402 & 21 \\
		\hline
	\end{tabular}
	\label{tab:6}
\end{table*}
\begin{table*}
	\centering
	\caption{Pearson Correlation}
	\begin{tabular}{lcccccc}
		\hline
		& \multicolumn{5}{c}{Pearson Correlation} \\
		\cline{2-6}
		& rice\_price & policy\_dummy & policy\_lag1 & policy\_lag2 & lag3 \\
		\hline
		rice\_price & 1.000 & 0.651 & 0.694 & 0.752 & 0.798 \\
		policy\_dummy & 0.651 & 1.000 & 0.894 & 0.791 & 0.686 \\
		policy\_lag1 & 0.694 & 0.894 & 1.000 & 0.884 & 0.767 \\
		policy\_lag2 & 0.752 & 0.791 & 0.884 & 1.000 & 0.868 \\
		lag3 & 0.798 & 0.686 & 0.767 & 0.868 & 1.000 \\
		\hline
	\end{tabular}
	\label{tab:7}
\end{table*}
\begin{table*}
	\centering
	\caption{Significance (one-tailed)}
	\begin{tabular}{lcccccc}
		\hline
		& \multicolumn{5}{c}{Significance (one-tailed)} \\
		\cline{2-6}
		& rice\_price & policy\_dummy & policy\_lag1 & policy\_lag2 & lag3 \\
		\hline
		rice\_price &  & 0.001 & 0.000 & 0.000 & 0.000 \\
		policy\_dummy & 0.001 &  & 0.000 & 0.000 & 0.000 \\
		policy\_lag1 & 0.000 & 0.000 &  & 0.000 & 0.000 \\
		policy\_lag2 & 0.000 & 0.000 & 0.000 &  & 0.000 \\
		lag3 & 0.000 & 0.000 & 0.000 & 0.000 &  \\
		\hline
	\end{tabular}
	\label{tab:8}
\end{table*}
\begin{table*}
	\centering
	\caption{Number of Cases}
	\begin{tabular}{lcccccc}
		\hline
		& \multicolumn{5}{c}{Number of Cases} \\
		\cline{2-6}
		& rice\_price & policy\_dummy & policy\_lag1 & policy\_lag2 & lag3 \\
		\hline
		rice\_price & 21 & 21 & 21 & 21 & 21 \\
		policy\_dummy & 21 & 21 & 21 & 21 & 21 \\
		policy\_lag1 & 21 & 21 & 21 & 21 & 21 \\
		policy\_lag2 & 21 & 21 & 21 & 21 & 21 \\
		lag3 & 21 & 21 & 21 & 21 & 21 \\
		\hline
	\end{tabular}
	\label{tab:9}
\end{table*}
\begin{table*}
	\centering
	\begin{minipage}[t]{0.37\textwidth} 
		\centering
		\caption{Coefficients - Unstandardized Coefficients}
		\setlength{\tabcolsep}{1pt} 
		\begin{tabular}{lcccc}
			\hline
			Variable & B & Std. Error & t & Significance \\
			\hline
			(Constant) & 13.346 & 3.138 & 4.253 & 0.001 \\
			policy\_dummy & 5.557 & 12.153 & 0.457 & 0.654 \\
			policy\_lag1 & 0.044 & 16.604 & 0.003 & 0.998 \\
			policy\_lag2 & 5.035 & 16.604 & 0.303 & 0.766 \\
			lag3 & 26.311 & 13.126 & 2.004 & 0.062 \\
			\hline
		\end{tabular}
		\label{tab:10}
	\end{minipage}
	\hfill 
	\begin{minipage}[t]{0.58\textwidth} 
		\centering
		\caption{Coefficients - Standardized Coefficients and Other Statistics}
		\setlength{\tabcolsep}{1pt} 
		\begin{tabular}{lccccccc}
			\hline
			Variable & $\beta$ & Correlations & Col Stats \\
			&  & Zero-order & Partial & Part & Tolerance & VIF \\
			\hline
			(Constant) &  &  &  &  &  &  \\
			policy\_dummy & 0.149 & 0.651 & 0.114 & 0.067 & 0.200 & 5.000 \\
			policy\_lag1 & 0.001 & 0.694 & 0.001 & 0.000 & 0.117 & 8.571 \\
			policy\_lag2 & 0.122 & 0.752 & 0.076 & 0.044 & 0.131 & 7.619 \\
			lag3 & 0.589 & 0.798 & 0.448 & 0.293 & 0.247 & 4.048 \\
			\hline
		\end{tabular}
		\label{tab:11}
	\end{minipage}
\end{table*}

\begin{figure}
	\centering
	\begin{subfigure}{0.45\textwidth}
		\centering
		\includegraphics[width=\textwidth]{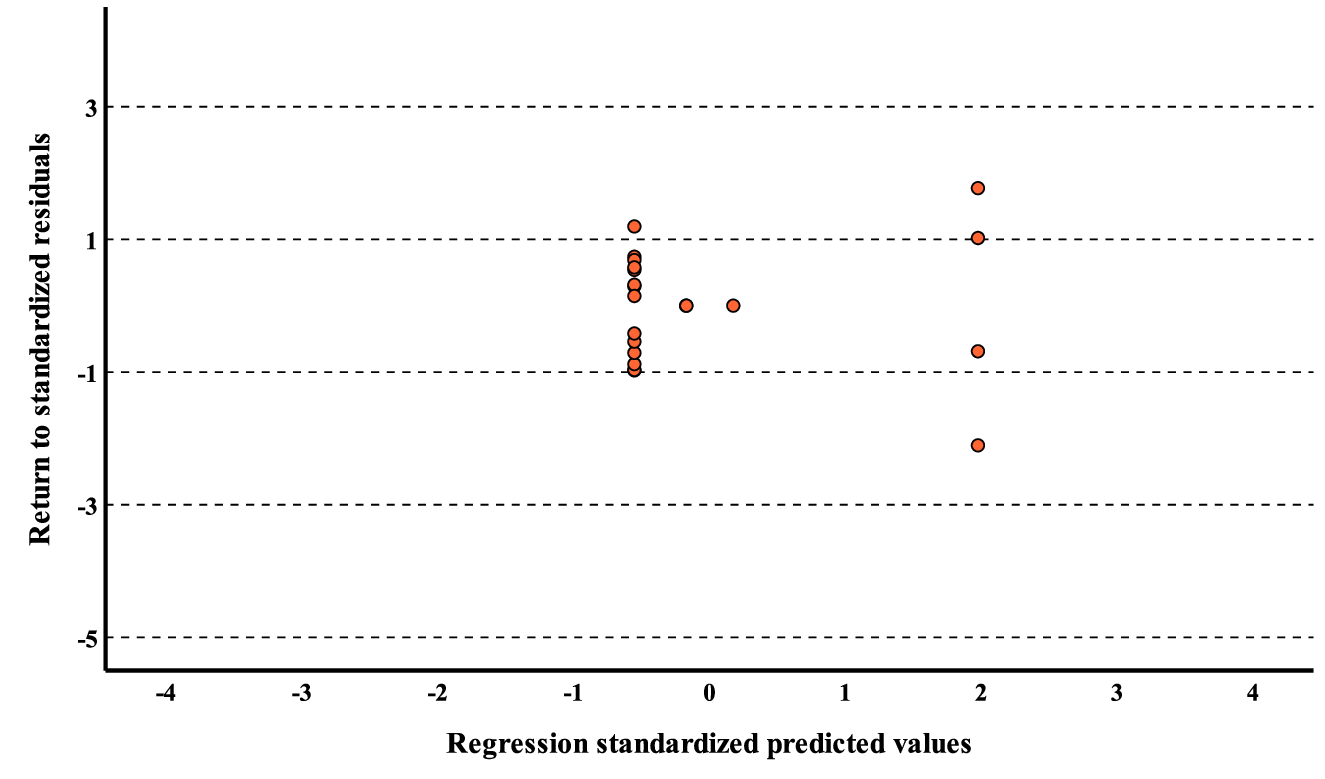}
		\caption{Residual statistics chart of the lagging regression model of the Old Taiwan dollar exchange policy impacting rice prices}
		\label{fig:4}
	\end{subfigure}
	
	\hfill 
	
	\begin{subfigure}{0.45\textwidth}
		\centering
		\includegraphics[width=\textwidth]{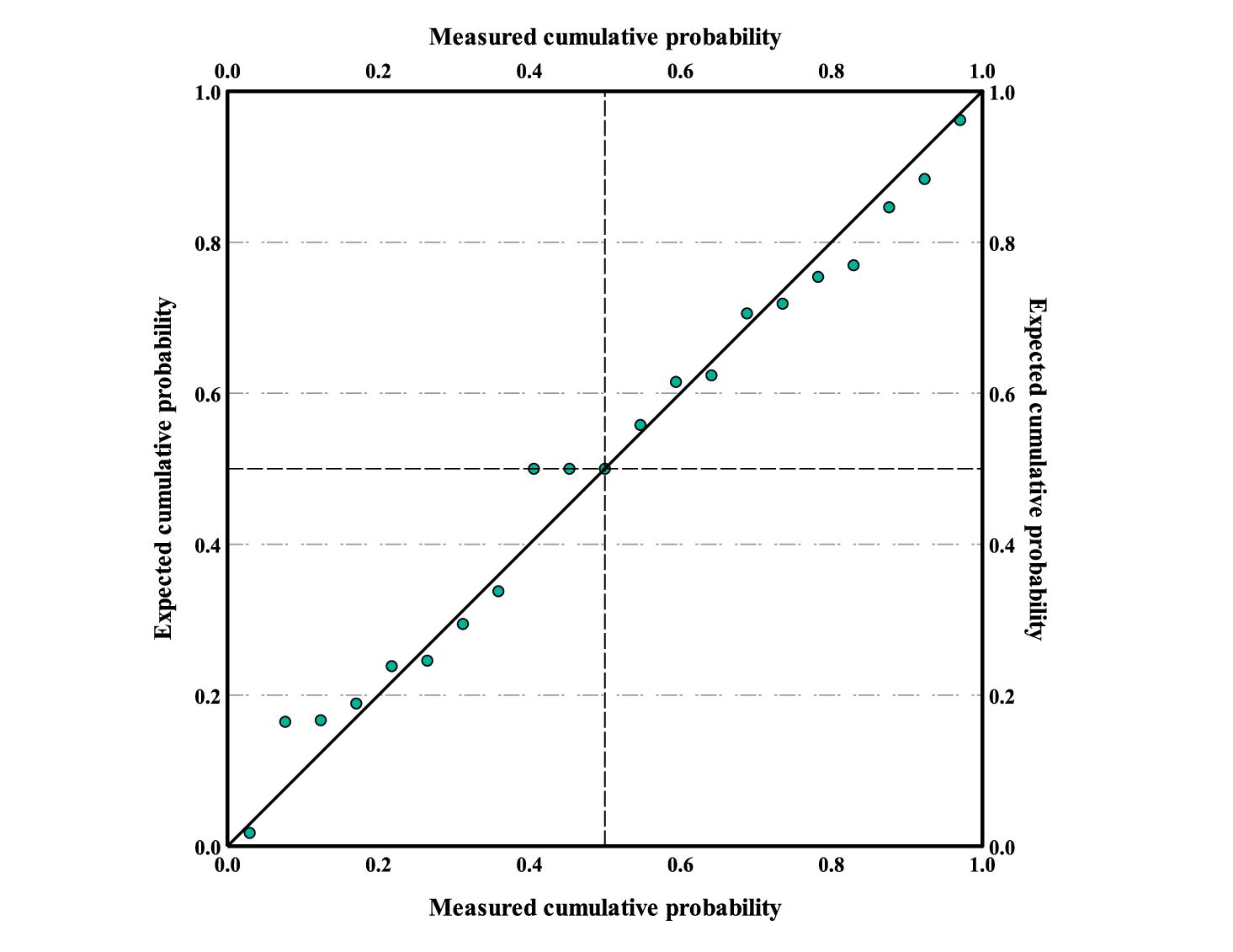}
		\caption{P-P plot}
		\label{fig:5}
	\end{subfigure}
	
	\hfill 
	
	\begin{subfigure}{0.45\textwidth}
		\centering
		\includegraphics[width=\textwidth]{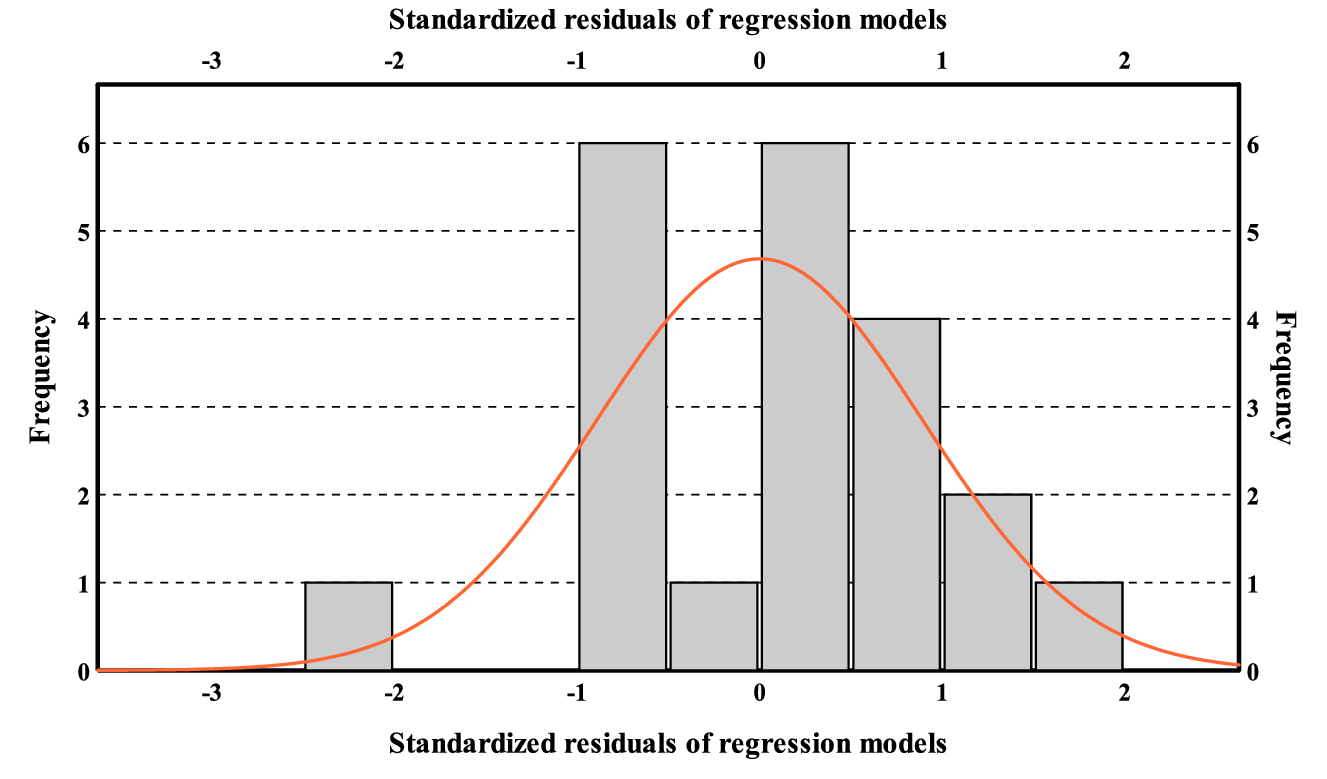}
		\caption{Standardized residual histogram}
		\label{fig:6}
	\end{subfigure}
	\caption{Residual analysis}
	\label{fig:figure3}
\end{figure}

As shown in Table.\ref{tab:6}, descriptive statistics are provided for the variables included in the regression analysis. The mean rice price is 21.4214 OTD/jin with a standard deviation of 17.9888. The policy dummy variable has a mean of 0.33, indicating that approximately 33 percents of the observations are in the post-policy period. The lagged policy variables (policy\textunderscore lag1, policy\textunderscore lag2, and lag3) also show varying degrees of correlation with rice prices.

As shown in Table.\ref{tab:7}, Pearson correlation coefficients are presented for the variables included in the regression analysis. The results indicate strong correlations between rice prices and the policy dummy variables (policy\textunderscore dummy, policy \textunderscore lag1, policy\textunderscore lag2, and lag3), with correlation coefficients ranging from 0.651 to 0.798.  These correlations suggest that policy changes have a significant association with rice price fluctuations.

As shown in Table.\ref{tab:8}, the significance levels for the Pearson correlation coefficients are presented.  All correlation coefficients are significant at the one-tailed level (p < 0.001), confirming the strong statistical relationship between rice prices and the policy variables.

As shown in Table.\ref{tab:9}, the number of cases for each variable included in the regression analysis is presented. The dataset includes 21 observations for each variable (rice\textunderscore price, policy\textunderscore dummy, policy\textunderscore lag1, policy\textunderscore lag2, and lag3), ensuring sufficient sample size for the analysis.

As shown in Table.\ref{tab:10}, the unstandardized coefficients from the regression analysis are presented. The results indicate that the policy dummy variable (policy\textunderscore dummy) has a coefficient of 5.557, suggesting a positive but statistically insignificant (p = 0.654) impact on rice prices.  The lagged policy variables (policy\textunderscore lag1, policy\textunderscore lag2, and lag3) also show varying degrees of impact, with lag3 having the highest coefficient (26.311) and approaching significance (p = 0.062).  

As shown in Table.\ref{tab:11}, the standardized coefficients and other statistics from the regression analysis are presented. The results indicate that the policy dummy variable has a standardized coefficient (Beta) of 0.149, while the lagged policy variables (policy\textunderscore lag1, policy\textunderscore lag2, and lag3) have Beta values ranging from 0.001 to 0.589.  The collinearity statistics (Tolerance and VIF) suggest potential multicollinearity issues, particularly for policy\textunderscore lag1 and policy\textunderscore lag2, which require further attention in future research. At present, the lag2 variable has marginal significance in the model, and lag has good significance.

We evaluated the performance of the regression model and the satisfaction of its assumptions by analyzing three key charts, as shown in Figure.\ref{fig:4}, Figure.\ref{fig:5} and Figure.\ref{fig:6} First, we examined the standardized residual histogram. This figure shows the distribution of the residuals of the regression model. Ideally, the residuals should approach a normal distribution, which is an important prerequisite for the effectiveness of linear regression models. The shape of the histogram in the figure is roughly bell-shaped and symmetrical to zero, which indicates that the residuals of the model are close to a normal distribution, thereby supporting the validity of the model. 

However, we also noticed that there is a certain deviation between the histogram and the normal distribution curve, especially at the tail end, which may suggest that there are some nonlinear relationships or omitted variables in the model. Furthermore, there are no obvious outliers in the histogram, which further enhances our confidence in the robustness of the model. Secondly, we analyzed the standardized predicted value graph. Although the specific content is not detailed, such charts are usually used to show the distribution of the model's predicted values. An ideal distribution of predicted values should be uniform and free of significant deviations, which indicates that the model's predictions are stable and accurate. If the predicted values have systematic biases or outliers, it may be necessary to further adjust the model or conduct in-depth analysis of the data features. Finally, we examined the measurement cumulative probability graph, which compares the cumulative probability distribution of the actual observed values with the model's predicted values. Ideally, the two should be close to a 45-degree diagonal, indicating that the model's prediction is highly consistent with the actual observed values. If the cumulative probability graph deviates significantly from the diagonal, it may indicate that there is a systematic error in the model's prediction. 

In this study, the cumulative probability graph is close to the diagonal, which indicates that the model's predictions are consistent with the actual observations, further verifying the validity of the model. To sum up, through the analysis of these three charts, we have gained a deeper understanding of the performance of the regression model. Although the model performs well in some aspects, there is still room for improvement, especially in handling nonlinear relationships and potential missing variables. Future research can further explore these aspects to enhance the predictive accuracy and explanatory power of the model.
\section{Conclusion}
\subsection{Four Stages of Rice Price Changes}
This study has provided a comprehensive quantitative reassessment of rice price dynamics during the 1945-1947 famine in post-war Taiwan.    By analyzing high-frequency rice price data and employing regression models, we identified four distinct stages of rice price changes and explored the correlations between these changes and government policies.

The first stage (August 1945-September 1945) was characterized by stable and slightly declining prices, reflecting the early post-war recovery period.
The second stage (September 1945-May 1946) saw a significant increase in rice prices, marking the first peak of the grain shortage.
The third stage (May 1946-November 1946) was a period of relative stability and a gradual decline in prices.
The fourth stage (November 1946-April 1947) witnessed sharp fluctuations and a sustained high level of rice prices, leading to social unrest.
\subsection{Policy Impact and Regression Analysis}
Our analysis revealed a significant correlation between rice prices and government policies, particularly the currency exchange policy implemented by the Chen Yi government.    The results of the independent samples t-test and effect size indicators (Cohen's d, Hedges' g, and Glass's $\Delta$) demonstrated that the policy had a substantial impact on rice prices, with significant differences observed before and after its implementation.

The regression models, particularly the exponential models used for the first two stages and the fourth stage, provided a good fit for the data, highlighting the exponential growth and decline patterns of rice prices during these periods. However, our analysis also identified potential multicollinearity issues, suggesting the need for further exploration of additional economic variables and more complex modeling techniques in future research.
\subsection{Contributions and Future Research Directions}
This study contributes to the understanding of the complex interplay between economic policies and food price dynamics in a historical context. The findings highlight the dominant role of policy systems in shaping post-war food markets and provide insights into the consequences of policy interventions during periods of economic instability. By quantitatively reassessing the rice price dynamics, this research offers a valuable reference for policymakers and scholars interested in the historical and contemporary dynamics of food markets and inflation.

While this study has provided significant insights, several avenues for future research remain open. Future studies could explore the impact of additional economic variables, such as inflation rates, exchange rates, and agricultural production levels, on rice prices. Moreover, extending the analysis to other regions or time periods could offer a broader perspective on the relationship between policy interventions and food price dynamics. Finally, the application of more advanced econometric techniques, such as panel data models or structural equation models, could enhance the robustness and accuracy of the findings. Of course, not considering other factors in the regression model is to simplify our modeling, to ignore all these factors into "constants", and to focus only on the policy of the Chen Yi government that we have selected. Besides, many studies have confirmed other factors.

In conclusion, this study underscores the critical role of policy in shaping food markets and highlights the importance of careful policy design and implementation to mitigate food shortages and price volatility. The findings serve as a reminder of the far-reaching consequences of economic policies on social stability and welfare.

\end{document}